\documentclass[aps,reprint,superscriptaddress,twocolumn,floatfix,showpacs,longbibliography]{revtex4-1}

\usepackage{graphicx,color}
\usepackage[utf8]{inputenc}
\usepackage{amsmath,amssymb,bm,amsfonts,dsfont,mathrsfs,amsthm}
\usepackage{braket}
\usepackage{txfonts,comment}
\usepackage[colorlinks=true,linkcolor=blue,citecolor=blue,urlcolor=blue]{hyperref}
\usepackage{soul}
\usepackage{enumitem}
\usepackage{multirow}
\usepackage{orcidlink}

\newcommand{\nn}{\nonumber \\}

\AtBeginDocument{%
    \newwrite\bibnotes
    \def\bibnotesext{Notes.bib}
    \immediate\openout\bibnotes=\jobname\bibnotesext
    \immediate\write\bibnotes{@CONTROL{REVTEX41Control}}
    \immediate\write\bibnotes{@CONTROL{%
    apsrev41Control,author="08",editor="1",pages="1",title="0",year="1"}}
     \if@filesw
     \immediate\write\@auxout{\string\citation{apsrev41Control}}%
    \fi
}%

\begin{document}

\title{Quantum-enhanced sensing of spin-orbit coupling without fine tuning}

\author{Bin Yi\,\orcidlink{0000-0002-3504-8173}}
\email{ucapbyi@uestc.edu.cn}
\affiliation{Institute of Fundamental and Frontier Sciences, University of Electronic Science and Technology of China, Chengdu 611731, China}
\affiliation{Key Laboratory of Quantum Physics and Photonic Quantum Information, Ministry of Education, University of Electronic Science and Technology of China, Chengdu 611731 , China}

\author{Abolfazl Bayat\,\orcidlink{0000-0003-3852-4558}}
\email{abolfazl.bayat@uestc.edu.cn}
\affiliation{Institute of Fundamental and Frontier Sciences, University of Electronic Science and Technology of China, Chengdu 611731, China}
\affiliation{Key Laboratory of Quantum Physics and Photonic Quantum Information, Ministry of Education, University of Electronic Science and Technology of China, Chengdu 611731 , China}
\affiliation{Shimmer Center, Tianfu Jiangxi Laboratory, Chengdu 641419, China}

\author{Saubhik Sarkar\,\orcidlink{0000-0002-2933-2792}}
\email{saubhik.sarkar@uestc.edu.cn}
\affiliation{Institute of Fundamental and Frontier Sciences, University of Electronic Science and Technology of China, Chengdu 611731, China}
\affiliation{Key Laboratory of Quantum Physics and Photonic Quantum Information, Ministry of Education, University of Electronic Science and Technology of China, Chengdu 611731 , China}

\begin{abstract}

Spin-orbit coupling plays an important role in both fundamental physics and technological applications. 
Precise estimation of the spin-orbit coupling is necessary for accurate designing across various physical setups such as solid state devices and quantum hardware. 
Here, we exploit quantum features in a 1D quantum wire for estimating the Rashba spin-orbit coupling with enhanced sensitivity beyond the capability of classical probes. 
The Heisenberg limited enhanced precision is achieved across a wide range of parameters and does not require fine tuning. 
Such advantage is directly related to the gap-closing nature of the probe across the entire relevant range of  parameters. 
This provides clear advantage over conventional criticality-based quantum sensors in which quantum enhanced sensitivity can only be achieved through fine-tuning around the phase transition point. 
We have demonstrated quantum enhanced sensitivity for both single particle and interacting many-body probes. 
In addition to extending our results to thermal states and the multi-parameter scenario, we have provided an measurement basis to perform close to the ultimate precision.

\end{abstract}

\maketitle

\section{Introduction}

The original notion of Spin-Orbit Coupling (SOC) refers to the relativistic quantum mechanical effect, connecting the spin and motional degrees of freedom of electrons in an electric field.
SOC plays a crucial role in solid state systems by affecting the energy spectrum in ways that has led to major research areas such as spintronics~\cite{nagaosa2010anamolous, hirohata2020review}, topological insulators~\cite{hasan2010topological, qi2011topological}, quantum dot arrays~\cite{golovach2004phonon, bulaev2008spinorbit, tanttu2019controlling, iijima2020gate} and quantum simulation with cold gases under artificial gauge fields~\cite{dalibard2011colloquium, goldman2014light, zhang2018spin}.
In particular, the Rashba type of SOC~\cite{manchon2015new}, which may occur naturally in a solid state material~\cite{min2006intrinsic} or can be engineered in quantum systems~\cite{zheng2015rashba, governale2002quantum, governale2002spin}, plays a crucial role in a wide range of technological applications. This includes spin field-effect transistors~\cite{datta1990electronic}, memory and logic devices~\cite{miron2011perpendicular, fert2013skyrmions}, spin-orbit torque devices~\cite{miron2011fast, garello2013symmetry}, spin filters and  pumps~\cite{koga2002rashba}, dissipation-less spin transport~\cite{sinova2015spin} and interaction-based quantum transport~\cite{kumar2021interactions}. 
In addition, Rashba SOC plays a significant role in emerging quantum technologies from solid state quantum simulators~\cite{reyren2007superconducting, caviglia2010tunable, brinkman2007magnetic, benshalom2010tuning} to fault-tolerant quantum computation through formation of Majorana fermions~\cite{alicea2012new} and information encoding in pseudo-spins~\cite{xu2014spin}.
Therefore, precise knowledge of the SOC parameters is necessary for studying both equilibrium and non-equilibrium properties of the system.
So far, SOC has been measured using various methods such as, electron transport~\cite{knap1996weak,luo1990effects}, spectroscopy~\cite{lashell1996spin, nitta1997gate}, and spin Hall effect~\cite{wunderlich2005experimental}.
The fundamental and technological importance of SOC and the recent developments in the field of quantum sensing pose a timely question: can the precision, with which the SOC parameters are measured, be enhanced using quantum features?

Quantum sensing has now established itself as a key component of quantum technologies that can surpass the limits of traditional classical sensors~\cite{degen2017quantum}.
The Cram\'{e}r-Rao inequality from estimation theory provides a lower bound on the precision of estimating an unknown parameter in terms of inverse of the Fisher information. 
Generally, as the probe size $L$ is increased, Fisher information grows as $ F {\sim }L^\beta $. 
For classical probes, the best achievable scaling is the standard limit where $ \beta {=} 1 $. 
Meanwhile, exploiting quantum features can improve the precision beyond the capacity of classical probes, with $ \beta {>} 1 $~\cite{paris2009quantum, degen2017quantum, braun2018quantum}, which is known as quantum-enhanced sensitivity.
The concept of quantum-enhanced sensitivity was first introduced in an interferometric setup utilizing Greenberger–Horne–Zeilinger (GHZ) type entangled states~\cite{giovannetti2004quantum, giovannetti2006quantum, giovannetti2011advances}. 
It significantly improves the precision of phase shift measurements to achieve the Heisenberg limit, $ \beta {=} 2 $. 
Since then, this phenomenon has been verified in various platforms including ion-traps~\cite{leibfried2004toward}, optical systems~\cite{mitchell2004super,pezze2007phase,ono2013entanglement}, superconducting qubits~\cite{wang2019heisenberg} and nitrogen-vacancy centers~\cite{bonato2016optimized}. 
However, the approach exclusively relies on GHZ-type entangled states, which are challenging to generate and highly susceptible to decoherence and particle loss. 
Aside from entanglement, squeezing is another established resource for interferometric phase estimation, especially with photonic setups~\cite{pezze2008mach, schnabel2017squeezed, polino2020photonic} and spin squeezed systems~\cite{ma2009fisher, frerot2018quantum, ma2011quantum}, see Ref.~\cite{maccone2020squeezing} for a unified framework of squeezing-based metrology.
These methods perform with a phase generating operation to encode information into the quantum state of the probe. 
Any disturbance in this unitary process reduces the precision, therefore limits the interferometric sensing method to scale up 
 effectively~\cite{de2013quantum}.

Another approach to quantum sensing utilizes many-body probes by leveraging a variety of quantum features. 
Although interactions have destructive effects in the GHZ-based sensing schemes, they play the primary role in many-body sensors~\cite{montenegro2025review}. 
In particular, quantum criticality has been identified as a resource for achieving quantum enhanced sensitivity. 
This includes various types of criticalities, such as first-order~\cite{raghunandan2018high, mirkhalaf2018supersensitive, yang2019engineering, heugel2019quantum, sarkar2025first}, second-order~\cite{zanardi2006ground, zanardi2007mixed, zanardi2008quantum, invernizzi2008optimal, gu2010fidelity, gammelmark2011phase, skotiniotis2015quantum, rams2018limits, chen2021effects,chu2021dynamic, liu2021experimental, montenegro2021global}, Floquet~\cite{mishra2021driving,mishra2022integrable}, time crystal~\cite{montenegro2023quantum, iemini2023floquet, yousefjani2024discrete, gribben2024quantum,shukla2024prethermal}, Stark~\cite{he2023stark, yousefjani2024nonlinearity,yousefjani2023long} and quasi-periodic~\cite{sahoo2024localization} localization, and topological~\cite{sarkar2022free, sarkar2024critical,mukhopadhyay2024modular} phase transitions. 
In all these criticalities, a common theme that emerges as the resource for quantum enhancement of the sensing capability is the closing of the energy gap at the phase transition point~\cite{montenegro2025review}. 
Nonetheless, the phase transition occurs only at a particular value of a Hamiltonian parameter which restricts the quantum advantage to a small region  around the critical point.
As one moves away from the criticality, the energy gap reopens and the quantum enhancement starts to disappear, typically falling back to the standard limit of sensing. 
This necessitates fine-tuning, making it primarily useful for local sensing where substantial prior knowledge about the parameter of interest is required.
Therefore, it is desirable to have a quantum system that feature the gap closing behavior over a wide range of the parameter to be estimated.

In this paper, we show that 1D quantum wires with SOC can provide a platform for Heisenberg precision sensing over a large area in parameter space. 
This finding provides a route for many-body quantum probe without exploiting only the vicinity of a critical point. 
Therefore, it opens up a way for Heisenberg precision sensing without fine-tuning. 
In this work we first consider the single parameter estimation scenario, where the Heisenberg scaling is shown for the cases of a single-particle probe, many-body interacting probe, and thermal probe. 
We then address the multi-parameter estimation case and finally present the optimal measurement basis.

\section{Overview: parameter estimation}
\label{Overview}

\subsection{Cram\'{e}r-Rao bound}
\label{CRB}

To estimate $d$ number of unknown parameters $\boldsymbol{\alpha} = (\alpha_1, \alpha_2, \ldots, \alpha_d)$, the first step is to encode them in  a quantum probe whose density operator $\rho_{\boldsymbol{\alpha}}$ depends on $\boldsymbol{\alpha}$. 
To determine these parameters, one needs to perform a measurement, which is represented by a set of Positive Operator-Valued Measure (POVM) operators $\{\Pi_k\}$ applied to the probe. 
The outcomes follow a classical probability distribution, where the probability of the $k$-th outcome is given by $p_k(\boldsymbol{\alpha}) = \mathrm{Tr}[\Pi_k \rho_{\boldsymbol{\alpha}}]$. 
These outcomes are fed into an estimator function to infer the parameters.
For this measurement setup, the uncertainty in estimating $\boldsymbol{\alpha}$ can be quantified by the covariance matrix $[\text{Cov}({\boldsymbol{\alpha}})]_{i,j} {=} \braket{\alpha_i \alpha_j} -\braket{\alpha_i} \braket{\alpha_j} $ .
Note that the diagonal elements are the single parameter variances $\sigma^2_{\alpha_i}$.
The lower bound in uncertainty is set by the Cram\'{e}r-Rao inequality~\cite{rao1992breakthroughs, braunstein1994statistical, cramer1999mathematical, liu2019quantum}
\begin{align}
\text{Cov}({\boldsymbol{\alpha}}) \geq \frac{1}{M} (\mathcal{F}^{C})^{-1} ,
\label{eq:CRB_c_multi}
\end{align}
where $M$ is the number of measurements and $\mathcal{F}^{C}$ is the $d \times d$ classical Fisher information matrix (CFIM) with elements
\begin{align}
\mathcal{F}^{C}_{i,j} = \sum_{k} \frac{\partial_{\alpha_i} p_k(\boldsymbol{\alpha}) \, \partial_{\alpha_j} p_k(\boldsymbol{\alpha})}{p_k(\boldsymbol{\alpha})} .
\label{eq:CFIM}
\end{align}
This basis-dependent lower bound can be optimized over all possible measurements to give the ultimate obtainable precision in terms of the quantum Cram\'{e}r-Rao bound~\cite{liu2019quantum}
\begin{align}
\text{Cov}({\boldsymbol{\alpha}}) \geq \frac{1}{M} (\mathcal{F}^{Q})^{-1} .
\label{eq:CRB_q_multi}
\end{align} 
The quantum Fisher information matrix (QFIM) elements can be expressed with the eigen-decomposition of the density operator $\rho_{\boldsymbol{\alpha}} {=} \sum_{n, (\lambda_n \ne 0)} \lambda_n \ket{\lambda_n} \bra{\lambda_n}$ in the following way~\cite{liu2019quantum},
\begin{align}
&\mathcal{F}^{Q}_{i,j} = \sum_{n, (\lambda_n \ne 0)} \frac{(\partial_{\alpha_i} \lambda_n) (\partial_{\alpha_j} \lambda_n)}{\lambda_n} \nn
&+ \sum_{n, (\lambda_n \ne 0)} 4 \lambda_n \text{Re}(\braket{\partial_{\alpha_i} \lambda_n | \partial_{\alpha_j} \lambda_n}) \nn 
&- \sum_{\substack{n,m \\ (\lambda_n, \lambda_m \ne 0)}} 8 \frac{\lambda_n \lambda_m}{\lambda_n+\lambda_m} \text{Re}(\braket{\partial_{\alpha_i} \lambda_n | \lambda_m} \braket{\lambda_m | \partial_{\alpha_j} \lambda_n}) .
\label{eq:QFIM_mixed}
\end{align}
For a pure state $\rho_{\boldsymbol{\alpha}} = \ket{\psi_{\boldsymbol{\alpha}}} \bra{\psi_{\boldsymbol{\alpha}}}$, QFIM becomes
\begin{align}
\mathcal{F}^{Q}_{i,j} = 4 \text{Re}(\braket{\partial_{\alpha_i} \psi | \partial_{\alpha_j} \psi}) - \braket{\partial_{\alpha_i} \psi | \psi} \braket{\psi | \partial_{\alpha_j} \psi}) .
\label{eq:QFIM_pure}
\end{align}

To extract a scalar form from the matrix inequalities, a positive weight matrix $W$ can be multiplied on both sides, followed by a trace operation. 
In particular, for the choice of $W$ being an identity matrix one gets $\sum_i \sigma^2_{\alpha_i} \geq \text{Tr}((\mathcal{F}^{Q})^{-1})/M$. Another useful choice is to take a $W$ matrix with all elements being zero except  $W_{i,i}{=}1$. 
Then the Cram\'{e}r-Rao inequality becomes $\sigma^2_{\alpha_i} \geq (\mathcal{F}^{Q})^{-1}_{i,i}/M \geq 1/ M \mathcal{F}^{Q}_{i,i}$, in which the last term is associated with the uncertainty in $\alpha_i$ when all the other parameters are known. 
This situation is known as the single parameter estimation problem where the error in estimating one unknown parameter $\alpha$ is lower bounded by $\sigma^2_{\alpha} \geq 1/M F^C(\alpha) \geq 1/ M F^Q(\alpha)$. 
Here, $F^C(\alpha)$ and $F^Q(\alpha)$ are the single parameter classical Fisher information (CFI) and quantum Fisher information (QFI), and are just the diagonal elements in CFIM and QFIM, respectively.
One choice of optimal basis, in which CFI saturates the QFI, is given by the projectors to the eigenstates of the so called symmetric logarithmic derivative operator (SLD) $\mathcal{L}$, which are implicitly defined for a probe state $\rho$ as $\partial_{\alpha} \rho {=} (\rho \mathcal{L} {+} \mathcal{L} \rho)/2$.

\subsection{Bayesian estimation}
\label{Bayesian}

The above description, known as the frequentist approach, expresses the lower bound of the precision in the case of ideal estimation.
A more practical and measurement data based approach is given by the Bayesian analysis.
In this approach, the parameters to be estimated ($\boldsymbol{\alpha}$) are treated as random variables.
An initial knowledge of the parameters is required in terms of a probability distribution, which is given by the \textit{prior} $P(\boldsymbol{\alpha})$.
If no prior information about the parameters are known except for their ranges, i.e.~$\alpha_i \in \left[\alpha_{i, \rm{min}},\alpha_{i, \rm{max}} \right]$, then the prior is a flat distribution $P(\boldsymbol{\alpha}) = \prod_i \frac{1}{\alpha_{i, \rm{max}} - \alpha_{i, \rm{min}}}$.
The experimental data is acquired by repeating the measurement process $M$ times where the $k$-th outcome is obtained $n_k$ times, with $\sum_k n_k {=} M$.
From this data, one can construct the \textit{likelihood} which is the conditional probability distribution $P(\{n_k\} | \boldsymbol{\alpha})$ for getting a dataset $\{n_k\}$ given that the parameters were $\boldsymbol{\alpha}$. 
This is given by the multinomial distribution, $P(\{n_k\} | \boldsymbol{\alpha}) = \frac{M !}{\prod_k n_k !} \prod_k p_k^{n_k}$, where $p_k$ is the theoretical probability of the $k$-th outcome.
The likelihood function compares the theoretical probability distribution $\{p_k\}$ with the experimentally obtained distribution $\{n_k/M\}$.
Now Bayes' theorem can be applied to write down the \textit{posterior} which is the conditional probability distribution for the parameters having values $\boldsymbol{\alpha}$ given the experimental outcomes,
\begin{align}
    P(\boldsymbol{\alpha} | \{n_k\}) = \frac{P(\{n_k\} | \boldsymbol{\alpha}) \, P(\boldsymbol{\alpha})}{\mathcal{N}} ,
\end{align}
where $\mathcal{N}$ is the normalization factor and is the probability of obtaining the experimental outcome.
This does not depend on $\boldsymbol{\alpha}$ and hence does not affect further analysis.
In the limit of large $M$, the central limit theorem applies, and the posterior takes a Gaussian form, the mean and variance of which give the estimated parameter values and precisions, respectively.

\subsection{Signal-to-noise ratio}
\label{SNR}

A popular and experimentally relevant tool to express the sensitivity of a sensing procedure is given by the signal-to-noise ratio (SNR), which can be written as $\alpha / \sigma_{\alpha}$ in a single parameter scenario~\cite{garbe2020critical}.
This ratio of the parameter to be estimated and the uncertainty, given by the precision achieved, quantifies the sensitivity.
In the optimal scenario, the Cram\'er-Rao bound can be saturated for large $M$ and thus SNR $\approx \alpha \sqrt{M F^Q(\alpha)}$.
The dependence on both the parameter value and the error makes SNR convey additional information than QFI.
Especially, in the context of weak field sensing,  where a much smaller error is required, the SNR value becomes particularly important.
In the multi-parameter case, this can be generalized~\cite{mihailescu2024multiparameter} to express the sensitivity for the $\alpha_i$ parameter as $\alpha_i \sqrt{M / (\mathcal{F}^{Q})^{-1}_{i,i}}$.
In this work, we showcase both these scenarios with the aid of this optimal SNR, as well as with the Bayesian case where the estimated value and precision is determined from measurement data.

\subsection{Numerical details}
\label{numerical}

Computation of QFI and CFI requires calculating the ground state and taking its partial derivative with respect to the parameter to be estimated.
In this work we use exact diagonalization to determine the ground state.
Even for the non-interacting many-body probes, the single-particle eigenstates can be utilized to express the QFI of the full ground state, as will be shown later.
For the interacting case, we are limited to size $L {<} 10$ for exactly computing the ground state.
To calculate the derivatives appearing in the expressions of QFI and CFI, we employ forward difference method to approximate $\ket{\partial_{\alpha} \psi} {\approx} \frac{\ket{\psi(\alpha+\Delta \alpha)} - \ket{\psi(\alpha)}}{\Delta \alpha}$.
We obtain the convergence for $\Delta \alpha$ typically in the range $10^{-4} - 10^{-3}$.
In our code we usually fix $\Delta \alpha {=} 10^{-6}$, as it is not numerically costly.
For the Bayesian analysis, the measurement outcome data is generated by uniform sampling from the ground states computed around the parameter to be estimated.

\section{Model: 1D quantum wire}
\label{Model}

The model considered in this work is based on the transport of electrons in a one-dimensional ballistic quantum wire in the presence of Rashba SOC~\cite{streda2003antisymmetric, birkholz2008spin}.
Low dimensional transport properties of electrons are of extreme importance in condensed matter physics, with applications in spintronics and quantum information processing.
Usually, such SOC is associated with two-dimensional electron gas systems in $x-y$ plane with strong confinement along the $z$-axis.
However, it is also prevalent to investigate interesting effects arising from motions in even reduced dimensions.
Further confinement to a wire geometry can enhance the effect of SOC~\cite{datta1990electronic} and has led to extensive studies on the transports of non-interacting electrons in quasi 1D systems.
The effective dynamics can be described by a tight-binding lattice model, which is governed by the Hamiltonian in the following form
\begin{equation}
H = H_0 + H_R + H_Z .
\label{eq:ham_main}
\end{equation}
Here $H_0$ describes the spin-independent nearest-neighbor hopping with strength $t$,
\begin{equation}
H_0 = - t \sum_{j,\sigma} (c^{\dag}_{j+1,\sigma} c_{j,\sigma} + c^{\dag}_{j,\sigma} c_{j+1,\sigma}),
\label{eq:ham0}
\end{equation}
where $c^{\dag}_{j,\sigma} (c_{j,\sigma})$ is the fermionic creation (annihilation) operator for site $j$ and spin $\sigma$ (denoting $\uparrow$ and $\downarrow$). 
The Rashba SOC term $H_R$ is
\begin{align}
H_R = & -\alpha_z \sum_{j,\sigma,\sigma'} (i\sigma_y)_{\sigma,\sigma'} c^{\dag}_{j+1,\sigma} c_{j,\sigma'} \nonumber \\
& + \alpha_y \sum_{j,\sigma,\sigma'} (i\sigma_z)_{\sigma,\sigma'} c^{\dag}_{j+1,\sigma} c_{j,\sigma'} + \text{H.c.} , 
\label{eq:hamR}
\end{align}
where $\alpha_{y(z)}$ are the SOC parameters and stem from the confining potentials of the wire. 
The Zeeman term $H_Z$ accounts for the presence of an external magnetic field $B$ along z direction,
\begin{align}
H_Z = B \sum_{j,\sigma,\sigma'}  (\sigma_z)_{\sigma,\sigma'} c^{\dag}_{j,\sigma}c_{j,\sigma'} . 
\label{eq:hamZ}
\end{align}

\section{Single Parameter Sensing}
\label{Single}

In low-dimensional transport experiments, accurate knowledge of the SOC parameters is desired.
Therefore, we first focus on estimating the SOC parameter in the uniform Rashba field scenario along $y$- and $z$-direction, i.e.~$\alpha_y {=} \alpha_z$.
The non-uniform SOC case will be addressed in the next section.
The goal is to analyze how well one can estimate $\alpha_z$ using the ground state of the system as a probe.
To stay close to the experimental systems, we consider finite wires with $L$ lattice sites and study the growth of QFI with system size. 
As the scaling of QFI is closely connected to the gap closing of the system, we also analyze the energy gap $\Delta$ between first excited state and ground state.
In our system, we also notice that the ground state is doubly degenerate, so a Zeeman term is essential to break the degeneracy.
This term need to be strong enough to capture the actual energy gap and not the Zeeman splitting.
For example, one should choose $B {>} 0.005t$ for $L {=} 100$.

\subsection{Single particle probe}
\label{Single particle}

\begin{figure}[t]
\centering
\includegraphics[width=0.47\textwidth]{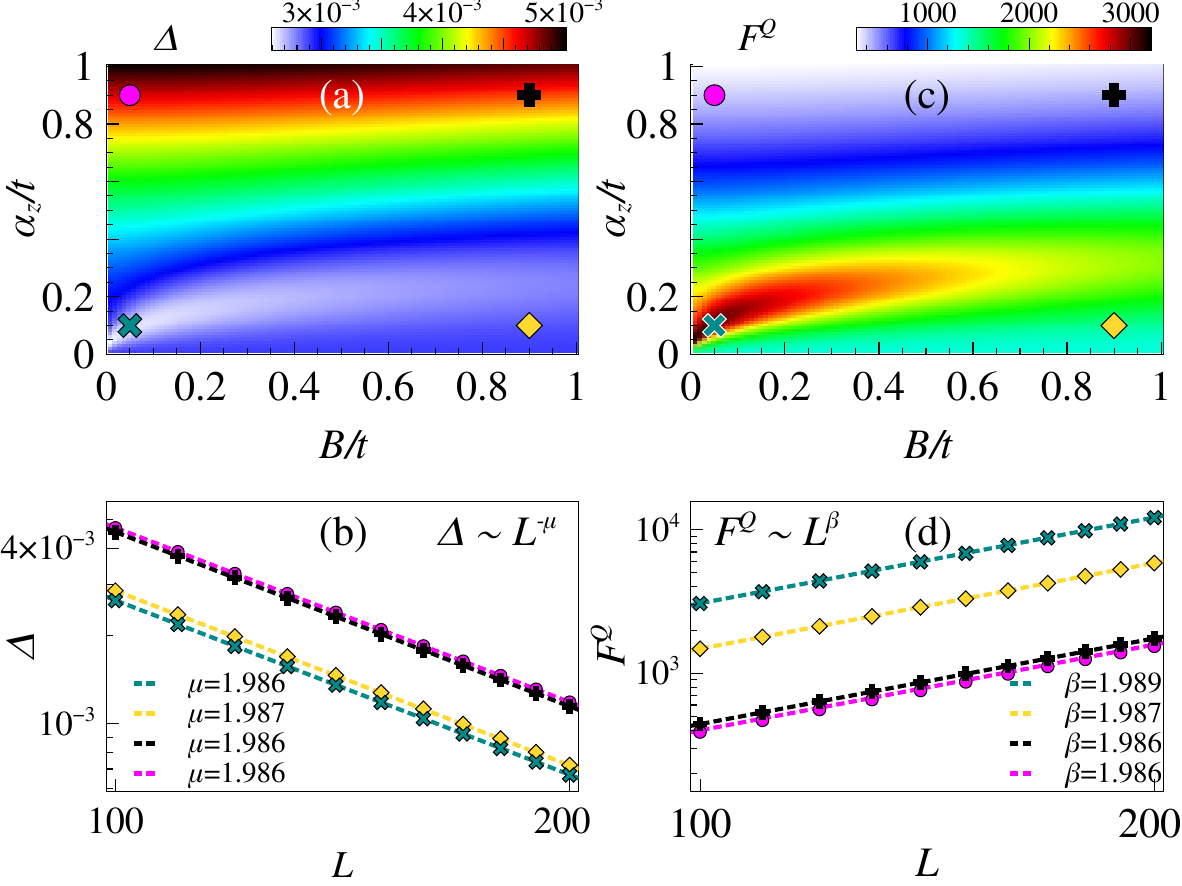}
\caption{\textbf{Scaling analysis of energy gap and QFI}. Probe state is the ground state of the Hamiltonian in Eqs.~\eqref{eq:ham_main}-\eqref{eq:hamZ} for sensing $\alpha_z$ with hopping parameter $t {=} 1$, and $\alpha_y {=} \alpha_z$. (a) Energy gap between first excited state and ground state $\Delta$ with $\alpha_z$ and $B$ for system size $L {=} 100$. The four points in the parameter space are chosen as $(B, \alpha_z) = (0.05, 0.1), (0.9, 0.1), (0.9, 0.9), (0.05, 0.9)$, denoted by the cross, diamond, plus, and circle, respectively. (b) Scaling relation of the energy gap with system size at these four parameter points, labeled by corresponding legends. In all cases numerical fit $\Delta {=} a L^{-\mu} {+} b$ shows almost quadratic scaling of the gap closing. (c) QFI of the probe with respect to $\alpha_z$ in the parameter space spanned by $\alpha_z$ and $B$ for system size $L {=} 100$. The same four points are chosen to study the scaling behavior. (d) Numerical fit confirms the quadratic scaling of QFI in the whole range as $F^{Q} {\sim} L^{\beta}$ with $\beta {\approx} 2$.}
\label{Fig_single}
\end{figure}

We start the sensing analysis with the single particle ground states of the Hamiltonian in Eq.\eqref{eq:ham_main} as the probe.
In Fig.~\ref{Fig_single}, we show the behavior of the energy gap $\Delta$ and QFI $F^{Q}$ of the ground state with respect to $\alpha_z$.
We have considered experimentally relevant values of $\alpha_z$ and $B$ for our analysis.
Fig.~\ref{Fig_single}(a) shows the result for a fixed size $L = 100$ where, for weak magnetic fields, the gap goes through a minimum at a small value of $\alpha_z$.
As the field becomes stronger, the minimum becomes less pronounced and the gap increases almost monotonically with $\alpha_z$.
Interestingly, the scaling of the decline of energy gap with system size shows similar behavior across the entire range of $\alpha_z$ and $B$.
This is shown in Fig.~\ref{Fig_single}(b), where four different combinations of $\alpha_z$ and $B$ values in the weak and strong regimes have been chosen to analyze the scaling.
In all the cases, the gap closing was found to scale almost quadratically with system size, i.e.~$\Delta \sim L^{-\mu}$ with $\mu \approx 2$. 
Indeed, for the QFI, Fig.~\ref{Fig_single}(c) shows almost a complementary picture to Fig.~\ref{Fig_single}(a), illustrating the close connection between the gap closing and the QFI.
This leads to a quadratic scaling of QFI in the whole region spanned by $\alpha_z$ and $B$, i.e.~$F^{Q} \sim L^{\beta}$ with $\beta \approx 2$, see Fig.~\ref{Fig_single}(d).
This is a very promising result as unlike criticality-based quantum sensors, where quantum-enhanced sensitivity is only achievable at the vicinity of the phase transition, one can reach Heisenberg scaling over a wide range of parameters. 
This can be explained by the favorable gap closing nature of the system.
We also note that for a given system size, the sensing capacity is stronger for smaller values of SOC parameters in the presence of a weak external magnetic field.
The measurement settings with which the error bounds given by the QFI can be approached is discussed later in Sec.~\ref{Optimal}.

\subsection{Many-body interacting probe}
\label{Interaction}

Now we study the effects of two-body contact interactions on the sensing capability of the probe. 
In one-dimensional systems, such  interactions are known to strongly influence the low-energy physics of many-body systems, leading to Luttinger liquid behavior.
The Hamiltonian in Eq.~\eqref{eq:ham_main} is modified to include the interaction between two fermions with different spins at the same site with strength $U$ and between two fermions on adjacent sites with strength $V$.
This results in rewriting the Hamiltonian as
\begin{equation}
H = H_0 + H_R + H_Z + H_{\rm int}, 
\label{eq:ham2}
\end{equation}
with,
\begin{equation}
H_{\rm int} = U \sum_{j} c^{\dag}_{j,\uparrow} c_{j,\downarrow} + V \sum_{j, \sigma, \sigma'} c^{\dag}_{j,\sigma}c_{j,\sigma} c^{\dag}_{j+1,\sigma'}c_{j+1,\sigma'} .
\label{eq:ham_int}
\end{equation}
The $U$ term accounts for an increase in the system's energy when two fermions with opposite spins occupy the same site.
For fermions in optical lattices, this term can be tuned by changing the scattering length with Feshbach resonance~\cite{chin2010feshbach}.
The $V$ term can arise due to the Coulomb repulsion, for example, between electrons confined in quantum dot arrays. 
As a typical scenario in theoretical and experimental situations, we consider the half-filled case here, i.e.~$L$ fermions on a $L$-site wire.
When $U {=} V {=} 0$, the fermionic ground state is given by the antisymmetric Slater determinant state formed by the single particle energy eigenstates $\{ \ket{E_n} \}$, which can be compactly written with the symmetric group $S_L$ as
\begin{align}
    \ket{\psi_{\rm GS}} = \frac{1}{\sqrt{L!}}\sum_{\boldsymbol{n} \in {S_L}}\text{sgn} (\boldsymbol{n}) \ket{E_{n_1}} \dots \ket{E_{n_L}}.
\end{align}
This results in an analytical QFI expression~\cite{sarkar2022free},
\begin{align}
    F^{Q} = 4\left(\sum_{l=1}^{L} \braket{\partial E_{l} | \partial E_{l}} - \sum_{l, l'} \braket{\partial E_{l} | E_{l'}} \braket{E_{l'} | \partial E_{l}} \right) .
\end{align}

\begin{figure}[t]
\centering
\includegraphics[width=0.48\textwidth]{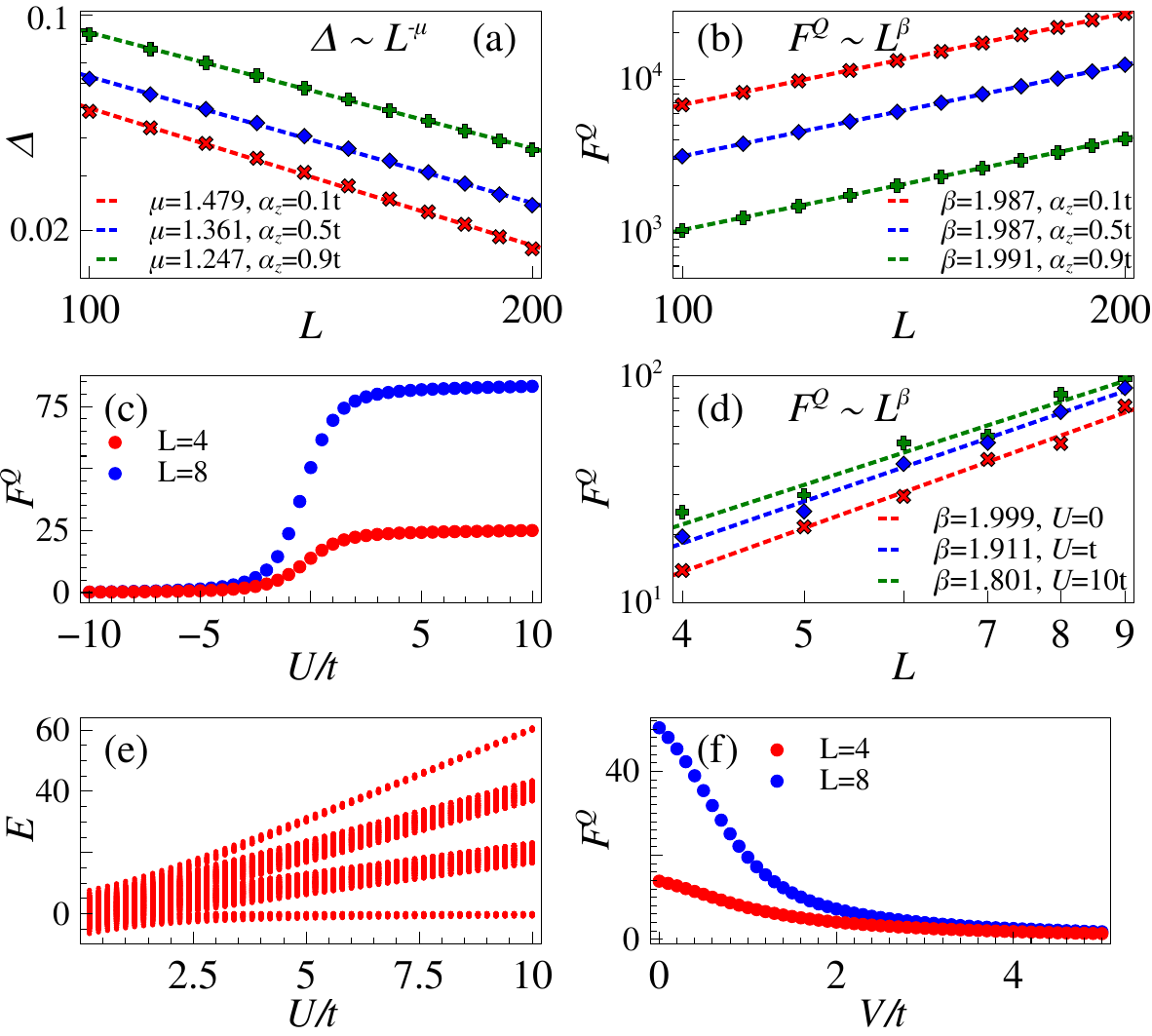}
\caption{\textbf{Many-body probe}. Probe state is the half-filled ground state of the Hamiltonian in Eq.~\eqref{eq:ham2} with $t {=} 1$, $B{=}0.01t$, and $\alpha_y {=} \alpha_z$. (a) Scaling of energy gap in the non-interacting case ($U{=} V {=} 0$) with system size $L$ for different values of SOC parameter $\alpha_z$. (b) Scaling of QFI with respect to $\alpha_z$ in the non-interacting case for different $\alpha_z$ values. (c) QFI at $\alpha_z {=} 0.1t$ as a function of $U$ when $V {=} 0$. (d) QFI scaling at $\alpha_z {=} 0.1t$ for different $U$ values. (e) Spectrum $E$ as a function of $U$. (f) QFI at $\alpha_z {=} 0.1t$ as a function of $V$ when $U {=} 0$.}
\label{Fig_many}
\end{figure}

Top row of Fig.~\ref{Fig_many} shows the results of the many-body non-interacting probe where the external field is fixed at $B {=} 0.01t$.
As the single particle eigenstates are filled up to $L$-th level in the ground state, the energy gap is defined by the energy difference between the $L{+}1$ and $L$-th eigenvalues.
As Fig.~\ref{Fig_many}(a) shows, gap closing is still algebraic with system size $(\Delta {\sim} L^{-\mu})$, but it is less than quadratic.
The exponent $\mu$ is found to be decreasing with increasing $\alpha_z$.
On the other hand, the QFI still scales quadratically with system size, as shown in Fig.~\ref{Fig_many}(b).
This shows that, the fermionic statistics in the non-interacting many-body ground state still results in the Heisenberg limit of precision for a wide range of the SOC parameter.
When only $U$ is non-zero, we observe that the QFI starts increasing with $U$ before attaining a saturation in the repulsive case.
For attractive $U$, the QFI decreases and goes to zero.
Fig.~\ref{Fig_many}(c) displays this behavior, while Fig.~\ref{Fig_many}(d) shows that algebraic scaling of QFI is still sustained in the repulsive case.
The exponential growth of the Hilbert space dimension limits our calculations in terms of system size, which can therefore be affected by finite size effects.
Nevertheless, the quadratic scaling of QFI for $U{=}0$ is expectedly observed in Fig.~\ref{Fig_many}(d).
Surprisingly, as $U$ is increased, the decrease in exponent $\beta$ is not drastic, which indicates that the interacting probes can still provide quantum enhancement.
This can be attributed to the fact that the presence of $U$ in Eq.~\eqref{eq:ham_int} does not open up the ground state energy gap as shown in the band structure in Fig.~\ref{Fig_many}(e). 
The fermions in the ground state stay localized at different sites.
In the other case, when $V$ is non-zero and $U {=} 0$, the QFI steadily decreases (see Fig.~\ref{Fig_many}(f)).
We numerically observe that quadratic scaling of QFI survives for small values of $V$ and disappears near $V {\approx} t$.

\subsection{Origin of quantum-enhancement}
\label{Origin}

In order to understand the underlying mechanism for the quantum-enhanced sensing with ground state probes we rely on recent analytical results of Ref.~\cite{abiuso2025fundamental}. According to these results,  the QFI of the ground state of a Hamiltonian $H_{\alpha} {=} H_1 {+} \alpha H_2$, is upper bounded by $||H_2||^2 / \Delta^2$, where $||{\bullet}||$ denotes the operator semi-norm (difference between highest and lowest eigenvalues)~\cite{abiuso2025fundamental}.
In our case, when considering the estimation of $\alpha_y {=} \alpha_z$, we have $H_2 {=} \sum_{j,\sigma,\sigma'} i(\sigma_z-\sigma_y)_{\sigma,\sigma'} c^{\dag}_{j+1,\sigma} c_{j,\sigma'} + \text{H.c.}$, and it can be checked quickly that $||H_2||$ does not scale with system size.
Therefore the enhancement solely comes form the scaling of the gap $\Delta {\sim} L^{-\mu}$ if $\mu {>} 1/2$.
While this only gives the upper bound, typically the same scaling behavior is displayed by the QFI near a phase transition~\cite{abiuso2025fundamental, sarkar2025first}.
Away from a transition, one needs to study on a case-by-case basis whether the quantum-enhancement is indeed manifested.
For the single-particle probe, we observe that the energy gap falls quadratically with system size (Fig.~\ref{Fig_single}(b)).
Therefore, we expect the QFI to show a quantum-enhanced behavior as its upper bound scales as $L^4$.
This shows that the observed quadratic scaling of QFI abides by the upper bound.
Although the quadratic scaling of the QFI is sustained in a wide range of the SOC strangth, the scaling advantage is lost when SOC becomes the dominating term in the Hamiltonian.
For example, when $\alpha_{z,y} \approx 10J$, the energy gap still falls as $L^{-2}$, but the ground state does not change appreciably with small changes in SOC strength.
Therefore, the QFI is close to zero and becomes size-independent for strong SOC.
For many-body non-interacting probe, the decrease of energy gap with system size occurs at a rate slower than quadratic fall-off (Fig.~\ref{Fig_many}(a)).
However, as the scaling is super-linear, we again expect the QFI to showcase the quantum enhancement.
As shown in Fig.~\ref{Fig_many}(b), we find that the QFI scales quadratically with system size, which is still within the upper bound.

Note that the resource with respect to which the scaling of the QFI is considered depends on the problem at hand.
Subject to the context, this can be particle number, system size, or time.
Although historically, Heisenberg limit has been obtained with respect to particle number~\cite{giovannetti2004quantum}, in a non-interacting ground state probe, it is possible to consider the system size instead as it determines the spatial extension over which the particle is delocalized.
The system size and particle number can even be interchangeable in some scenarios such as a spin-system that can be mapped on to spin-less fermions on a lattice with Jordan-Wigner transformation.
Although it is not applicable to our case, in a dynamics based probe, time serves as a resource for quantum enhancement~\cite{boixo2007generalized, ilias2022criticality}.
For our model, it is possible to look additionally at the scaling with respect to particle numbers.
However, it would lead to only linear scaling as the ground state would be a tensor product of fermionic single-particle eigenstates and the QFI is additive~\cite{he2023stark, manshouri2025quantum}.

\subsection{Thermal probe}
\label{Thermal}

\begin{figure}[t]
\centering
\includegraphics[width=0.49\textwidth]{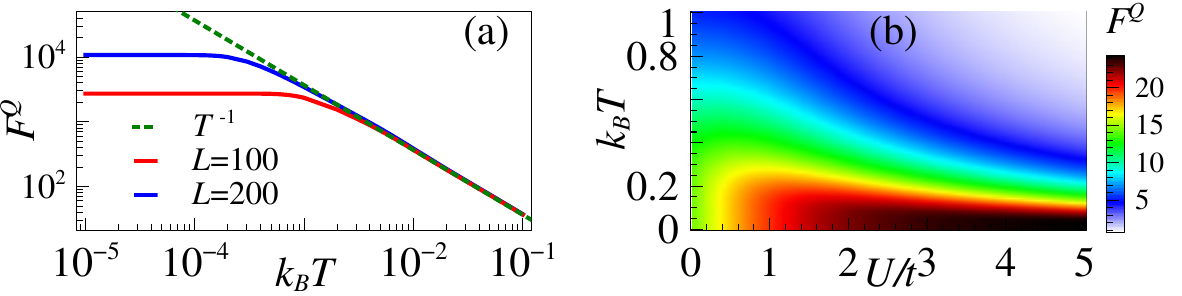}
\caption{\textbf{Thermal probe}. Probe state is the thermal state of the Hamiltonian in Eq.~\eqref{eq:ham2} with $t {=} 1$, $B{=}0.01t$, $\alpha_y {=} \alpha_z$. (a) QFI vs.~temperature in the non-interacting case for different system sizes. As thermal energy surpasses the gap, QFI falls of universally as ${\sim} 1/T$. (b) Variation of QFI with both $U$ and $T$ for $L {=}6 $. }
\label{Fig_temp}
\end{figure}

So far we have discussed the cases where the probe is taken to be the ground state which is the equilibrium state at zero temperature.
In practice however, the sensors generally operate at finite temperature and hence, it is necessary to analyze the effect of temperature on the sensing capability.
When the temperature is smaller than the energy gap i.e.~$k_B T {<} \Delta$ ($k_B$ is the Boltzmann constant), we expect the ground state description to hold. 
Higher temperatures result in a mixed state with increasing contributions from higher energy states. 
The QFI then starts to decrease and typically collapses universally with an algebraic decay with temperature~\cite{he2023stark}.
In Fig.~\ref{Fig_temp}(a), we show this behavior in the non-interacting case for two system sizes $L=100, 200$. 
The ground state QFI value is retained up to a temperature $T{\sim} \Delta/k_B$, after which the QFI is found to fall off as $T^{-1}$.
In the repulsive interacting case, we observe that the energy gap decreases with $U$, and hence the temperature at which the QFI starts to fall decreases with $U$ (see Fig.~\ref{Fig_temp}(b)).

\subsection{Disorder}
\label{disorder}

\begin{figure}[t]
\centering
\includegraphics[width=0.45\textwidth]{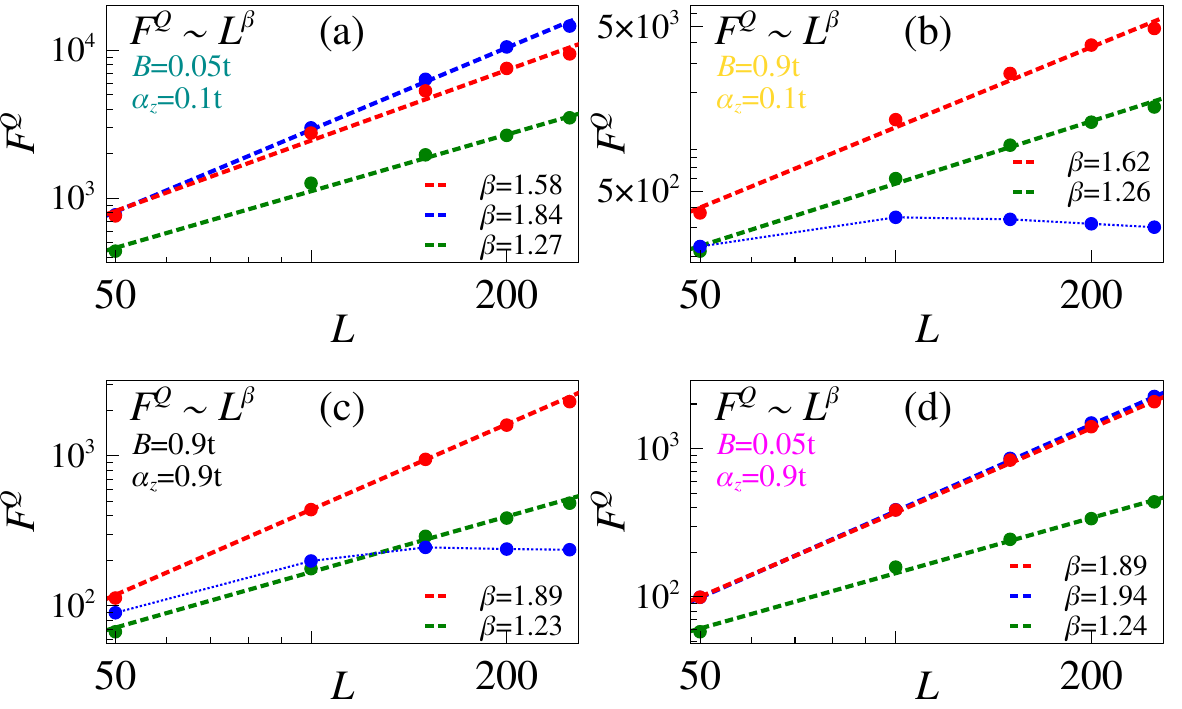}
\caption{\textbf{Disorder for single-particle probes}. $(B, \alpha_z) = (0.05, 0.1), (0.9, 0.1), (0.9, 0.9), (0.05, 0.9)$ in (a), (b), (c), and (d), respectively. The red, blue, and green circles represent disorder in tunneling, magnetic field, and onsite potential, respectively.}
\label{Fig_dis1}
\end{figure}

\begin{figure}[b]
\centering
\includegraphics[width=0.45\textwidth]{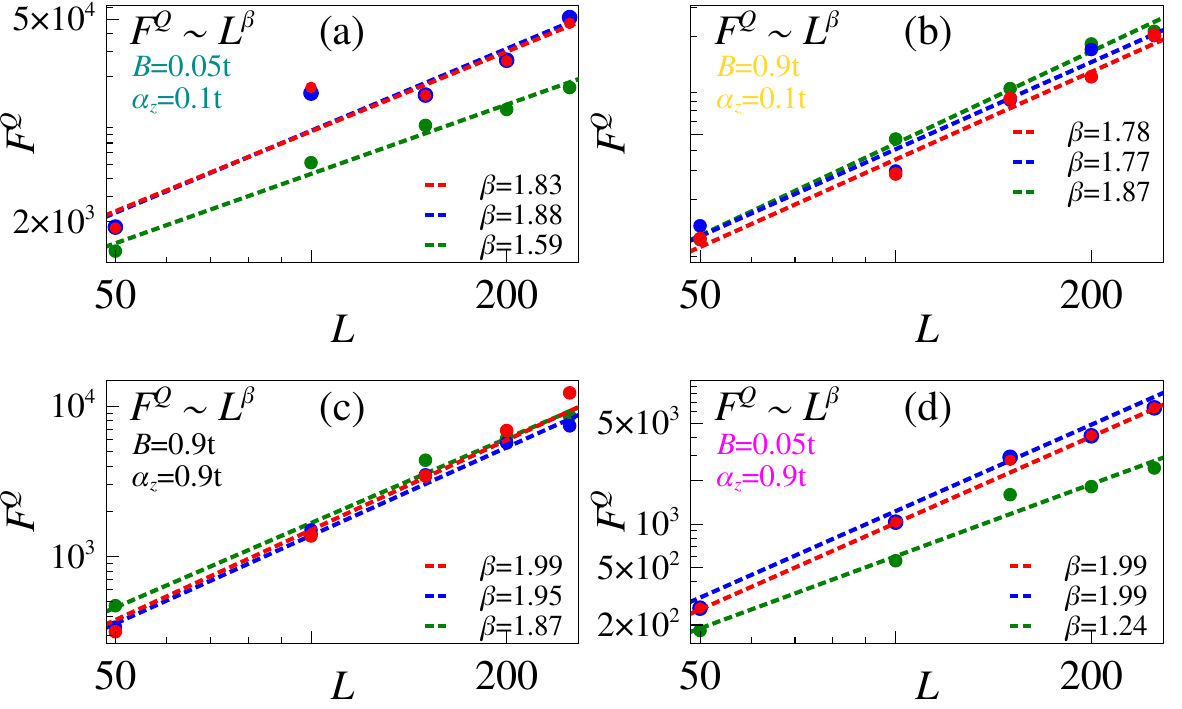}
\caption{\textbf{Disorder for many-body probes}. $(B, \alpha_z) = (0.05, 0.1), (0.9, 0.1), (0.9, 0.9), (0.05, 0.9)$ in (a), (b), (c), and (d), respectively. The red, blue, and green circles represent disorder in tunneling, magnetic field, and onsite potential, respectively.}
\label{Fig_dis2}
\end{figure}

After analyzing the effect of thermal noise, we now consider another practical issue that can take place in the form of disorders during physical realization.
The disorder can occur in various terms, including in the tunneling strength $t$ and in the applied magnetic field $B$, which is crucial for breaking the degeneracy.
Moreover lattice defects can also take place in the form of holes or unoccupied sites.
To see the effect of these disorders on the sensing, we include an additional term to the Hamiltonian in Eq.~\eqref{eq:ham_main}
\begin{align}
&H_{\rm dis} =  - \sum_{j,\sigma} \Delta t_j (c^{\dag}_{j+1,\sigma} c_{j,\sigma} + c^{\dag}_{j,\sigma} c_{j+1,\sigma}) \nn
&+ \sum_{j,\sigma,\sigma'} \Delta B_j \; (\sigma_z)_{\sigma,\sigma'} \; c^{\dag}_{j,\sigma}c_{j,\sigma'} 
+ \sum_{j, \sigma} \mu_j c^{\dag}_{j,\sigma}c_{j,\sigma}. 
\label{eq:ham_dis}
\end{align}
Here $\Delta t_j$ is a random disturbance to the tunneling between sites $j$ and $j{+}1$ that is drawn from a uniform distribution $[0, W_t]$, with $W_t$ as the disorder strength.
$\Delta B_j$ is a random magnetic field along $z$-direction at the $j$-th site that is drawn from a uniform distribution $[-W_B, W_B]$, with $W_B$ as the disorder strength.
Finally, $\mu_j$ is an onsite potential that takes a high value $\mu$ on some randomly chosen sites, thus making those sites practically unoccupied due to high energy penalties.
We examine the effect of each type of disorder individually with $W_t {=} 0.01 t$, $W_B {=} 0.1 B$, and $\mu {=} 10t$ with number density of holes as $L/50$.
We find these values to be typical disorder strengths and hole density up to which quantum enhancement is observed.
The departure from quadratic scaling due to disorder is shown in Fig.~\ref{Fig_dis1} for the single particle probe.
The four panels Figs.~\ref{Fig_dis1}(a), (b), (c), (d) correspond to the four points in the $B-\alpha_z$ plane that have been considered before as well.
In each panel, the scaling behavior is shown for the individual cases of disorder in $t$, $B$, and presence of holes.
Apart from the decrease in scaling exponent, we notice that enhancement vanishes for higher $B$ values as $W_B$ is too large (Figs.~\ref{Fig_dis1}(b) and (c)). 
The many-body non-interacting probes are found to be more robust against disorder, as shown in Fig.~\ref{Fig_dis2} for the same four points in the $B-\alpha_z$ plane.
Here we notice higher values of scaling exponent as well as enhancement even for higher $W_B$ values (Figs.~\ref{Fig_dis2}(b), (c)).
Therefore our analysis shows that even in the presence of imperfections, the sensing performance can show reduced advantage, which is encouraging from an experimental point of view.

\section{Multi-parameter sensing}
\label{Multi}

\begin{figure}[t]
\centering
\includegraphics[width=0.49\textwidth]{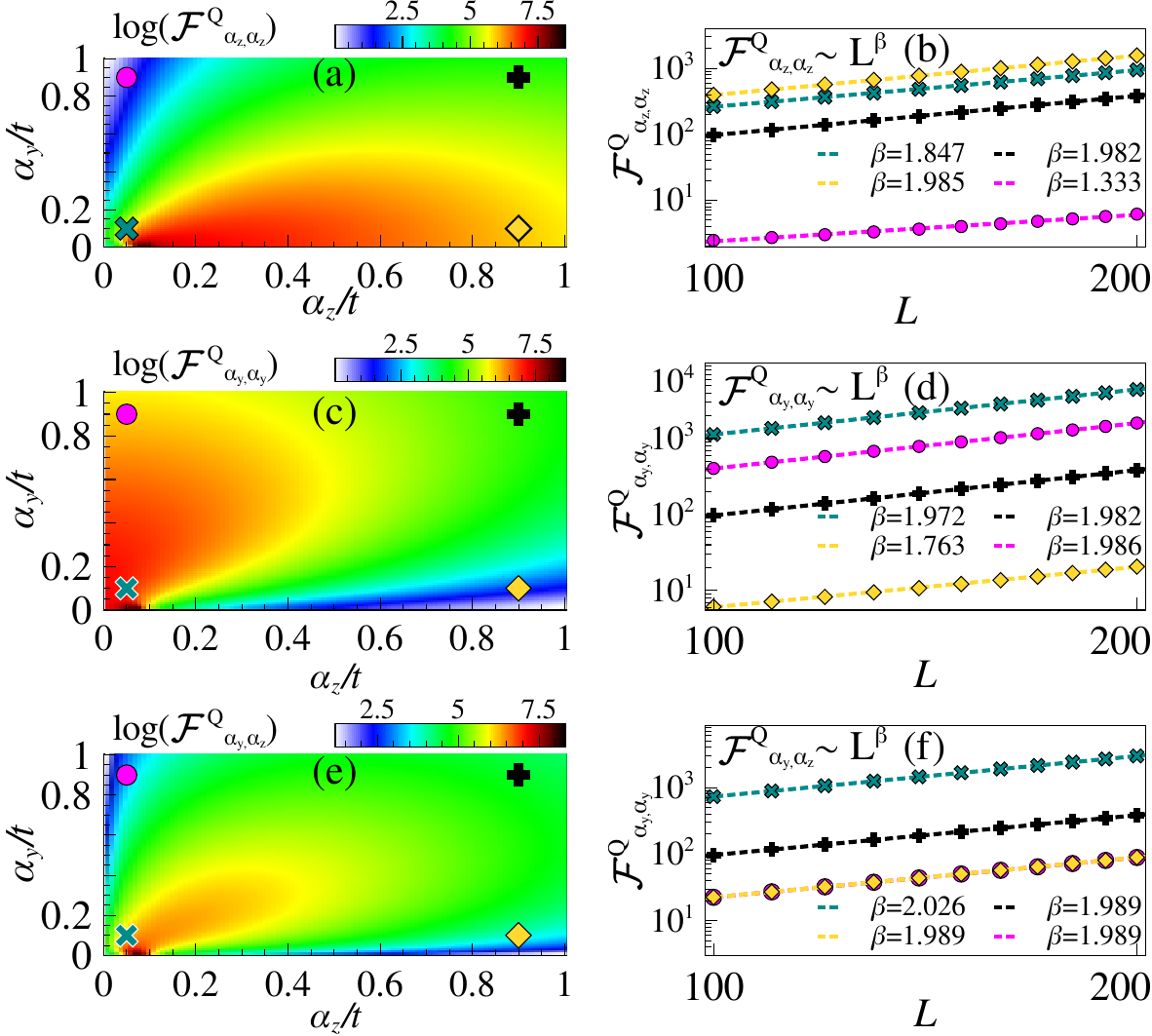}
\caption{\textbf{Multi-parameter sensing}. Probe state is the single particle ground state of the Hamiltonian in Eq.~\eqref{eq:ham_main} with $t {=} 1$, and $B{=}0.01t$. The QFIM components are computed using Eq.~\eqref{eq:QFIM_pure}. (a) The $(\alpha_z, \alpha_z)$ component of the QFIM $\mathcal{F}^Q$ across a range of $\alpha_y$ and $\alpha_z$ for $L{=}100$. Four points chosen in the parameter space for scaling analysis are $(\alpha_z, \alpha_y) = (0.05, 0.1), (0.9, 0.1), (0.9, 0.9), (0.05, 0.9)$, denoted by the cross, diamond, plus, and circle, respectively. (b) Scaling of $(\mathcal{F}^Q)_{\alpha_z, \alpha_z}$ at the four points. (c) The $(\alpha_y, \alpha_y)$ component of $\mathcal{F}^Q$. (d) Scaling of $(\mathcal{F}^Q)_{\alpha_y, \alpha_y}$ at the four points. (e) The $(\alpha_y, \alpha_z)$ component of $\mathcal{F}^Q$. (d) Scaling of $(\mathcal{F}^Q)_{\alpha_y, \alpha_z}$ at the four points. }
\label{Fig_multi}
\end{figure}

In this section, we consider a more general scenario where we do not presume prior knowledge of the SOC parameters i.e.~the Rashba fields in any direction. 
The goal is to estimate both $\alpha_y$ and $\alpha_z$ simultaneously, making the problem multi-parametric.
To analyze the system's sensitivity to changes in these parameters, we compute the multi-parameter QFIM $\mathcal{F}^{Q}$, which provides a comprehensive description of the ultimate precision limits for estimating multiple parameters concurrently. 
In Fig.~\ref{Fig_multi}(a) we shows the component $\mathcal{F}^{Q}_{\alpha_z, \alpha_z}$ as a function of the two parameters. 
Fig.~\ref{Fig_multi}(b) shows that $\mathcal{F}^{Q}_{\alpha_z, \alpha_z}$ reaches super-linear scaling in a wide range of $\alpha_z$ and $\alpha_y$ and can also attain close to Heisenberg scaling. 
Similar observation also holds for the component $\mathcal{F}^{Q}_{\alpha_y, \alpha_y}$, as shown in Figs.~\ref{Fig_multi}(c) and (d). From this analysis we can also infer that the quantum enhancement of the sensing of $\alpha_y$ when $\alpha_z$ is known, can reach the Heisenberg limit as well.
In general, however, the sensitivity depends on the off-diagonal component $\mathcal{F}^{Q}_{\alpha_y, \alpha_z}$ that captures the correlation between the estimates of $\alpha_z$ and $\alpha_y$. 
Figs.~\ref{Fig_multi}(e) and (f) show that the scaling of $\mathcal{F}^{Q}_{\alpha_y, \alpha_z}$ also stays close to quadratic across the whole range.
In the following, we address how these QFI bounds can be obtained with a practical measurement basis.

\section{Optimal measurement}
\label{Optimal}

\begin{figure}[t]
\centering
\includegraphics[width=0.49\textwidth]{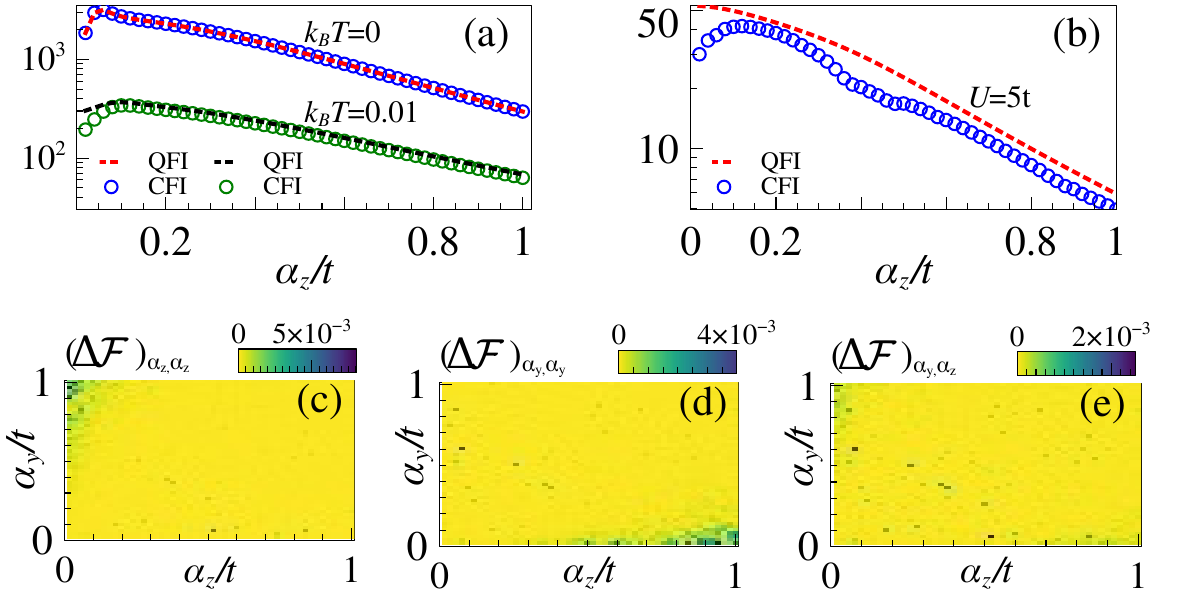}
\caption{\textbf{Comparison between QFI and CFI}. The CFI is computed in the eigenbasis of the current operator defined in Eq.~\eqref{eq:current}. (Top row) The dotted lines and the circles denote the QFI and CFI, respectively. (a) For a single-particle probe with $L {=} 100$, the top curves correspond to ground state and bottom curves correspond to thermal state at $k_{B}T {=} 0.01t$. (b) Many-body interacting probe with $L {=} 6$ and $U {=} 5t$.
(Bottom row) The relative difference $\Delta \mathcal{F}$ between QFIM and CFIM across a range of $\alpha_y$ and $\alpha_z$ for $L{=}100$ and $B{=}0.01t$. 
(c) The $(\alpha_z, \alpha_z)$ component of $\Delta \mathcal{F}$.
(d) The $(\alpha_y, \alpha_y)$ component of $\Delta \mathcal{F}$.
(e) The $(\alpha_y, \alpha_z)$ component of $\Delta \mathcal{F}$.}
\label{Fig_CFI}
\end{figure}

While the QFI gives the ultimate bound for precision, the optimal measurement basis is not unique.
Although it is mathematically possible to find such a basis by diagonalizing the aforementioned SLD operator, this procedure usually leads to a complicated and non-local measurement choice.
In practice, it is therefore advisable to look for an experimentally implementable measurement basis with which the calculated CFI is quite close to the QFI.
In our work, we find that the lattice version of the particle current operator~\cite{hetenyi2009many, dutta2012integer}, defined as 
\begin{equation}
\mathcal{I} = i \sum_{j,\sigma} (c^{\dag}_{j+1,\sigma} c_{j,\sigma} - c^{\dag}_{j,\sigma} c_{j+1,\sigma}), 
\label{eq:current}
\end{equation}
generates a suitable basis for estimation. 
The CFI measured in the eigenbasis of $\mathcal{I}$ matches with the corresponding QFI for single-particle probe, as shown in Fig.~\ref{Fig_CFI}(a). 
In fact, by carrying out a Fourier transform to the momentum-space, we can readily see that $\mathcal{I} {=} \sum_k 2\sin{(k) \, c^{\dag}_{k,\sigma} c_{k,\sigma}}$.
Therefore, the eigenbasis is the simple momentum basis which can be measured with existing experimental techniques such as the time-of-flight method.
It is also possible to measure such current operators in optical lattice experiments~\cite{impertro2024local, impertro2025strongly}.
Thermal energy exceeding the gap and interaction results in small discrepancy between the CFI and QFI for small $\alpha_z$ , as shown in Figs.~\ref{Fig_CFI} (a) and (b), respectively.

The multi-parameter sensing scenario is more challenging because here the different optimal measurement bases for different parameters do not necessarily commute with each other~\cite{ragy2016compatibility} and one also needs to make sure that the matrices are not singular~\cite{mihailescu2025metrological}.
Fortunately, for the model considered here, the current operator $\mathcal{I}$ provides the optimal basis for estimating both parameters as the CFIM $\mathcal{F}^{C}$ (defined in Eq.~\eqref{eq:CFIM}) calculated in this basis matches quite well with the QFIM.
We define the relative difference between the QFIM and CFIM by $(\Delta \mathcal{F})_{i,j} {=} (\mathcal{F}^{Q} {-} \mathcal{F}^{C})_{i,j} / \mathcal{F}^{Q}_{i,j}$.
In Figs.~\ref{Fig_CFI} (c-e), we show that different elements of the $\Delta \mathcal{F}$ matrix are indeed very close to zero across a wide range of $\alpha_z$ and $\alpha_y$.

\section{Precision analysis}
\label{precision}

\begin{figure}[t]
\centering
\includegraphics[width=0.49\textwidth]{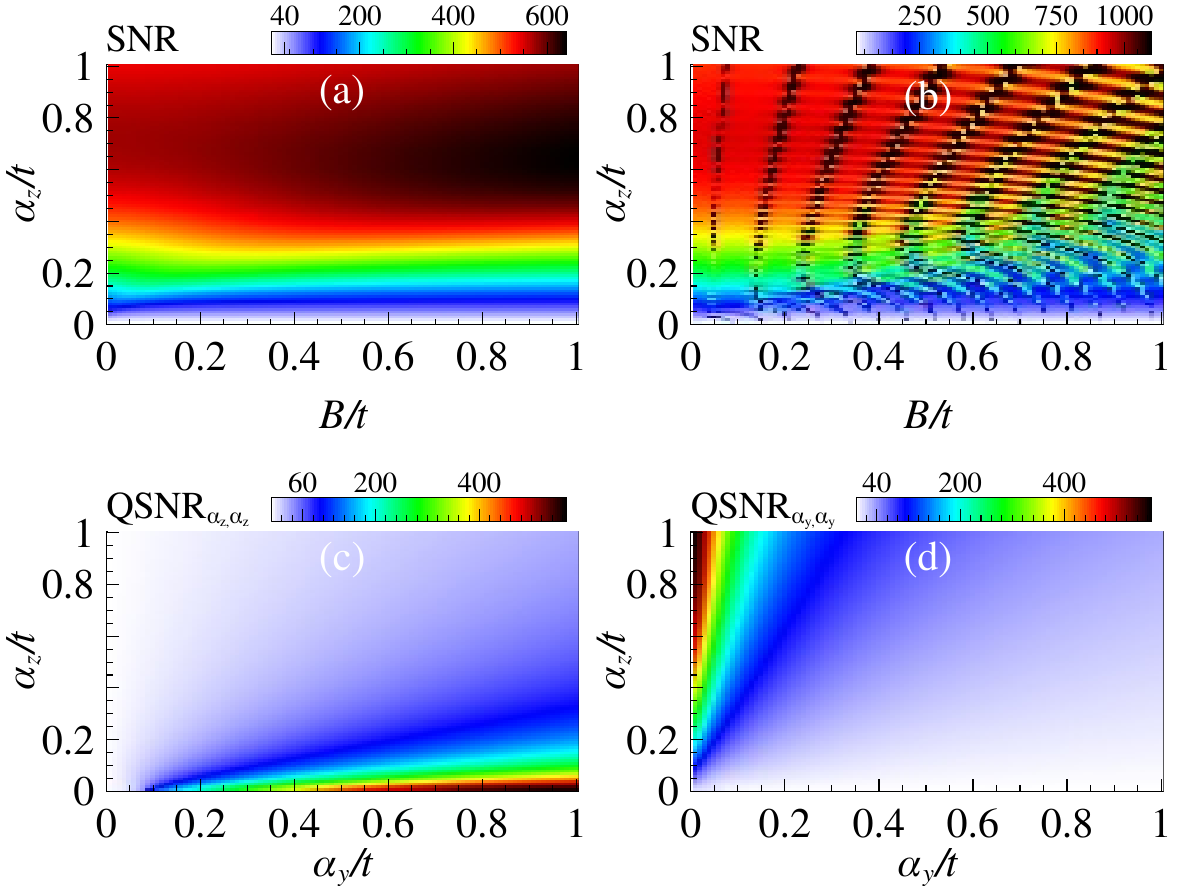}
\caption{\textbf{Sensitivity from optimal SNR} for system size $L {=} 100$. (a) Single-particle probe. (b) Many-body non-interacting probe with half-filling. Multi-parameter scenario with precision given by QFIM elements (c) $(\mathcal{F}^Q)_{\alpha_z, \alpha_z}$  and (d) $(\mathcal{F}^Q)_{\alpha_y, \alpha_y}$.}
\label{Fig_SNR}
\end{figure}

We now examine the precision obtained for the various probes considered in this work.
The favorable scaling behavior of QFI without needing fine-tuning is quite encouraging, but we can also notice that the value of QFI itself can depend on the parameters themselves. 
Since the QFI exhibits a nearly consistent super-linear scaling $\sim L^\beta$ across a broad parameter range, its dependence on the unknown parameter comes from the prefactor.
This can be seen, for example, in Figs.~\ref{Fig_single}(d),~\ref{Fig_many}(b), and in Fig.~\ref{Fig_multi}.
It is quite evident from Fig.~\ref{Fig_single}(c), that for a given $B$ value, the maximum QFI occurs at a certain $\alpha_z$.
Therefore, it is crucial to analyze the sensitivity across the ranges of the parameters to examine the merit of the metrological advantage.
To this effect, we report the SNR in Fig.~\ref{Fig_SNR} which shows the ratio of the parameter and the precision achievable in the optimal case.  
As mentioned in Sec.~\ref{SNR}, this is given by $\alpha_z \sqrt{M F^Q(\alpha_z)}$ for the single-parameter case, and by $\alpha_i \sqrt{M / (\mathcal{F}^{Q})^{-1}_{i,i}}$ for the multi-parameter case.
We observe that, even with experimentally low-cost value measurement repetition number $M {=} 1000$, the SNR values obtained are quite high, as SNR in the range $10 - 100$ is already considered satisfactory in physical realizations.
This is shown for single-particle probes in Fig.~\ref{Fig_SNR}(a), for many-body non-interacting probes in Fig.~\ref{Fig_SNR}(b), and for the multi-parameter scenario in Figs.~\ref{Fig_SNR}(c) and (d).
As is evident from the Figs.~\ref{Fig_SNR}(a)-(b), the SNR values are lower for small $\alpha_z$ values, but the estimation errors are still at least one order of magnitude smaller than the parameter values themselves.
This relays the importance of the SNR analysis.
We also notice that the SNR values create certain arrays of local maxima for the many-body case (Fig.~\ref{Fig_SNR}(b)), for specific combinations of $\alpha_z$ and $B$ values.
This is caused by the overall complicated spectrum of the filled energy levels that results in local minimum in energy gap structure, which subsequently increases the QFI at those points.

\begin{figure}[t]
\centering
\includegraphics[width=0.49\textwidth]{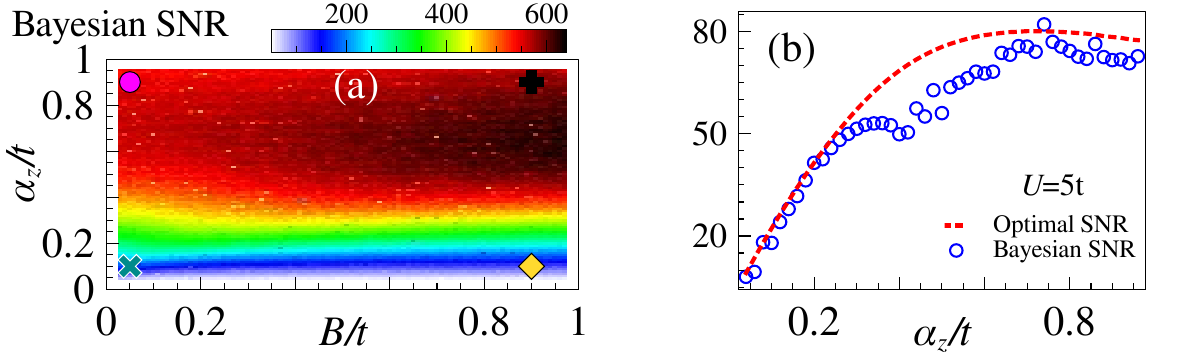}
\caption{\textbf{Sensitivity from Bayesian analysis}. (a) Single-particle probe with $L {=} 100$. (b) Many-body interacting probe with $L {=} 6$ and $U {=} 5t$.}
\label{Fig_Bayesian}
\end{figure}

Beyond optimal estimation, one can carry out similar analysis by using more practical estimation strategies such as Bayesian approaches.
We show these results in Fig.~\ref{Fig_Bayesian} by quantifying the sensitivity as the ratio of the estimated parameter and the achieved precision for the same number of measurement repetitions as before.
In this case, we utilize the basis laid out in Sec.~\ref{Optimal} to generate the measurement data.
The measurement data is generated by sampling from the probability distribution provided by the probe state in this basis.
The posterior distribution obtained, as outlined in Sec.~\ref{Bayesian}, provides the error that quantifies the sensitivity.
Again we observe quite satisfactory values for the sensitivity obtained for the single particle probe in Fig.~\ref{Fig_Bayesian}(a) and a many-body interacting probe in Fig.~\ref{Fig_Bayesian}(b).
Therefore we can infer that the performances of the different probes are quite competent, along with the advantages coming from the scaling properties and the lack of the requirement of fine-tuning.

\section{Conclusion}
\label{Conclusion}

Due to the long reaching impact of SOC on various fields ranging from condensed matter systems to emerging quantum technologies,  it is of utmost importance to estimate the SOC strength with high precision. 
In this paper, we employ  tools from quantum sensing to achieve quantum enhanced sensitivity for the estimation of Rashba SOC in one dimensional quantum wires. 
This enhancement is closely connected to gap closing  feature of the system.
Unlike conventional criticality based quantum sensors, in which the energy gap closing happens at the phase transition point, our probe achieves quantum advantage with Heisenberg precision across a wide range. 
Our analysis with both single-particle and many-body interacting probes establishes quantum advantage for all such parameters within characteristic temperature range.
We have extended our results to multi-parameter sensing scheme where multiple SOC terms can be jointly estimated. 
In addition, we show that a simple particle current measurement can closely reach the ultimate precision limit.
This work can be extended to search for other models where enhanced sensitivity can be achieved over a wide range.
Parallelly, one can also extend the study for other types of SOC and even other types sensors such as field sensors.
Another avenue worth exploring is the possibility to utilize non-equilibrium dynamics for estimating SOC as they are experimentally more stratightforward to employ.

\begin{acknowledgments}

SS acknowledges support from National Natural Science Foundation of China (Grant No.~W2433012).
AB acknowledges support from the National Natural Science Foundation of China (grants No. 12274059, No. 12574528, No. 1251101297 and No. W2541020).

\end{acknowledgments}

\section*{Data availability}
The data that support the findings of this article are openly available~\cite{githublink}.

\bibliography{Ref}

\end{document}